\documentclass[conference]{IEEEtran}
\IEEEoverridecommandlockouts
\usepackage{amsmath,amsfonts}
\usepackage{algorithmic}
\usepackage{algorithm}
\usepackage{array}
\usepackage[caption=false,font=normalsize,labelfont=sf,textfont=sf]{subfig}
\usepackage{textcomp}
\usepackage{stfloats}
\usepackage{url}
\usepackage{verbatim}
\usepackage{graphicx}
\usepackage{caption}
\usepackage{romannum}
\usepackage{cite}
\usepackage{amsmath}
\usepackage{makecell}
\usepackage{multirow}
\usepackage{hyperref}
\usepackage{makecell}
\usepackage{xcolor
}
\usepackage{adjustbox}
\begin{document}

\title{On the Potential of Digital Twins for Distribution System State Estimation with Randomly Missing  Data in Heterogeneous Measurements
\\

% \author{Yuanshuo Zhang,
%         Ying Zhang,~\IEEEmembership{Member,~IEEE},}
%         % <-this % stops a space
\vspace{-5pt}
\thanks{This research was supported (in part) by the National Science Foundation through award number 2418359.}
}
% \vspace{-10pt}

\author{
\IEEEauthorblockN{Ying Zhang {}}
\IEEEauthorblockA{\textit{School of Electrical and Computer Engineering} \\
\textit{Oklahoma State University},
Stillwater, OK, U.S. \\
y.zhang@okstate.edu}
\and

\IEEEauthorblockN{Yihao Wang}
\IEEEauthorblockA{
\textit{Department of Computer Science} \\
\textit{Southern Methodist University},
Dallas, TX, U.S. \\
yhwang@smu.edu}
\and 

\IEEEauthorblockN{Yuanshuo Zhang}
\IEEEauthorblockA{\textit{School of Electrical and Computer Engineering} \\
\textit{Oklahoma State University},
Stillwater, OK, U.S. \\
yuanshuo.zhang@okstate.edu}
\and
\IEEEauthorblockN{Eric Larson}
\IEEEauthorblockA{
\textit{Department of Computer Science} \\
\textit{Southern Methodist University},
Dallas, TX, U.S. \\
eclarson@lyle.smu.edu}

\and
\IEEEauthorblockN{Di Shi}
\IEEEauthorblockA{\textit{School of Electrical and Computer Engineering} \\
\textit{New Mexico State University}, Las Cruces, NM, U.S. \\
dshi@nmsu.edu}

\and
\IEEEauthorblockN{Fanping Sui}
\IEEEauthorblockA{
\textit{Department of Mechanical Engineering}\\
\textit{University of California Berkeley}, Berkeley, CA, U.S. \\
fpsui@berkeley.edu}
}

% \author{
% \IEEEauthorblockN{Ying Zhang*}
% \IEEEauthorblockA{\textit{School of Electrical and Computer Engineering} \\
% Oklahoma State University,
% Stillwater, U.S. \\
% y.zhang@okstate.edu}
% \and

% \IEEEauthorblockN{Yihao Wang, Eric Larson}
% \IEEEauthorblockA{
% \textit{Department of Computer Science} \\
% Southern Methodist University,
% Dallas, U.S. \\
% yhwang@smu.edu; eclarson@lyle.smu.edu}
% \and

% \IEEEauthorblockN{Yuanshuo Zhang}
% \IEEEauthorblockA{School of Electrical and Computer Engineering \\
% Oklahoma State University,
% Stillwater, U.S. \\
% yuanshuo.zhang@okstate.edu}
% \and
%\and
% \IEEEauthorblockN{Fanping Sui}
% \IEEEauthorblockA{
% \textit{Department of Mechanical Engineering}\\
% University of California Berkeley, CA, U.S. \\
% fpsui@berkeley.edu}
% \and
% \IEEEauthorblockN{Di Shi}
% \IEEEauthorblockA{School of Electrical and Computer Engineering \\
% New Mexico State University, Las Cruces, U.S. \\
% dshi@nmsu.edu}
% 
% }
% \vspace{-15pt}
% The paper headers
% \thanks{
% % Y. Zhang is with the Department of Electrical and Computer Engineering, Oklahoma State University, Stillwater, OK 74074, USA (e-mail: y.zhang@okstate.edu).
% % Y. Wang is with the Department of Computer Science, Southern Methodist University, Dallas, TX 75205, USA (e-mail: yhwang@smu.edu).
% 
% }% <-this % stops a space
\maketitle

\begin{abstract}
Traditional statistical optimization-based state estimation (DSSE) algorithms rely on detailed grid parameters and mathematical assumptions of all possible uncertainties. Furthermore, random data missing due to communication failures, congestion, and cyberattacks, makes these methods easily infeasible. 
Inspired by recent advances in digital twins (DTs), this paper proposes an interactive attention-based DSSE model for robust grid monitoring by integrating three core components: physical entities, virtual modeling, and data fusion.  
To enable robustness against various data missing in heterogeneous measurements, we first propose physics-informed data augmentation and transfer. Moreover, a state-of-the-art attention-based spatiotemporal feature learning is proposed, followed by a novel cross-interaction feature fusion for robust voltage estimation. 
A case study in a real-world unbalanced 84-bus distribution system with raw data validates the accuracy and robustness of the proposed DT model in estimating voltage states, with random locational, arbitrary ratios (up to 40\% of total measurements) of data missing.
\end{abstract}

\vspace{-5pt}
\section{Introduction}

\IEEEPARstart{D}{igital} twins (DTs) represent a physical asset, system, or process in a digital counterpart, which can mirror the real-world power grids in its operating behaviors by utilizing data-based simulations
\cite{grieves2014digital}. The development
of advanced metering infrastructure, together with the Internet of Things (IoT) and artificial intelligence (AI) technologies, have made DT technology feasible and economical for industrial-level implementation \cite{DT10666925}.
% The digital twin analyzes the large volumes of data collected by metering infrastructure to obtain usable insights for sound decision-making.
Substantial efforts have been devoted to employing DTs in various grid applications, such as wind farm modeling and power transformer monitoring \cite{Zhang2023, moutis2020digital, menezes2024survey}.
% \cite{Zhang2023} constructs a digital twin of a wind turbine
% by employing physics models and the latest physics-informed deep learning techniques. 

The white paper on DTs \cite{grieves2014digital} defines the three main parts of a digital twin: physical entities in real space, virtual models in virtual space, and data that connects physical entities and virtual models together. Unlike conventional data-driven methods that focus solely on virtual modeling, this paper thus proposes the \textit{first-of-the-kind} DT model for distribution system state estimation (DSSE) with randomly missing data, by integrating the three parts in this dedicated application. State estimation is the backbone of power system operation, which attains real-time system-wide status by synthesizing heterogeneous measurement data. However, in some
distribution networks with poor observability, even the distribution system operators (DSOs) cannot maintain an accurate model of the network \cite{Huang10114954}. Data-driven DSSE techniques are thus considered effective solutions; however, they require large volumes of complete datasets without missing data, which constrains their practicality in real-world distribution grids. 

Missing data is quite common in field measurements due to sensor malfunctions, environmental interferences, communication disruptions, etc. \cite{10155244, 8440475}. 
% Also, data collection strategies might intentionally exclude certain measurements to reduce storage costs \cite{10155244}.
To alleviate the adverse effects of missing data, recent efforts \cite{10309251,8440475,9770510} develop statistical optimization-based state estimation methods, such as weighted least square (WLS) and Kalman filters. For example, 
\cite{10309251} introduced extra random binary variables that obey the Bernoulli process to represent if measurements are lost to the DSSE formulation. 
% \cite{9770510} adopts projection statistics for missing data and updates the noise covariance based on a Sage-Husa noise estimator.
However, most of these model-based methods rely on detailed grid parameters and mathematical assumptions of uncertainties of measurement errors. Moreover, they require prior statistical information about the missing data (e.g., the probability and location) and perform poorly when these assumptions are violated. 
% \cite{10155244} applies the interval $[0,1]$ to depict the missing data and designs the fault-tolerant extended Kalman filter for the recursive filter framework. 
% \cite{8440475} modifies the unscented Kalman filtering (KF) by modeling the arrival rate of missing and delayed measurements as a stochastic process. 
% \cite{deshmukh2013state} considers the impact of missing data by analyzing its effect on the stability of the nonlinear SE process, using an extended KF approach to handle missing data. 
% The missing data is addressed by incorporating the phenomenon of data dropouts in the design of the state estimator in \cite{7166299}.
% The recent widespread deployment of smart meters and supervisory control and data acquisition (SCADA) systems has led to a surge in data availability. 
% With available data,
Recently, data-driven imputation methods \cite{9639949,NTU7913716, raghuvamsi2024distribution} have attracted considerable attention. 
% However, challenges such as missing data can arise due to hardware failures and communication disruptions.
%If there are errors in the topology, such as unknown or incorrectly assessed states of certain switching devices, the SE may inaccurately infer the operational state and power flow within the grid. 
% To ensure the accuracy of state estimation with missing data, \cite{zargar2020probabilistic} proposes an optimal method for arranging PMUs to address the issue of missing data at certain nodes due to limited measurements, thereby ensuring the accuracy of SE calculations.
% In \cite{9639949}, deep neural networks (NNs) are proposed to perform supervised learning for DSSE with missing input data. 
% For example, \cite{NTU7913716} proposes ensemble data analytics on incomplete phasor measurement unit (PMU) measurements to assess power system stability. 
For DSSE, a transformer model integrated with a bi-directional long short-term memory (LSTM) layer is used in \cite{raghuvamsi2024distribution} to predict missing measurement data.
% \cite{raghuvamsi2024distribution1} proposes a convolutional generative adversarial network with modified loss functions. 
Nonetheless, these methods completely ignore multi-modal uncertainties of heterogeneous measurements and combine multimodal data into a single vector as the neural network (NN) input. Hence, only focusing on the virtual modeling, these data-driven methods often suffer from over-fitting and physical inconsistency issues, especially in the case of extensive missing data.
% They essentially overlook the differences in magnitudes between different physical quantities \cite{modality-multivar, modality-dualbranch}. 

To fill the gap, we propose an interactive attention-based DT model for DSSE with random missing data. 
% This model is featured by the flexibility and scalability of varying-length data input by performing NN-based spatiotemporal inference.
To handle
data missing scenarios, we first propose data augmentation and transfer for the DT modeling. Informed by physics, the proposed model then extracts feature representations for different physical quantities in a parallel manner, thereby efficiently reducing the computational burdens. Moreover, utilizing the acquirement of attention-based feature learning, a novel cross-interaction feature fusion algorithm is proposed. 
% The proposed algorithm consists of a state-of-the-art attention-based feature extraction and a novel cross-interaction feature fusion algorithm for voltage estimation.   
% and conduct cross-interaction feature fusion for state estimation with heterogeneous measurements. 
The proposed DT method exhibits salient features distinguished from the conventional data-driven method, including physical consistency and predictive capability for multi-modal uncertainties. Compared with \cite{9639949,raghuvamsi2024distribution}, the proposed method is end-to-end, without the separate data imputation, by integrating feature extraction, fusion, and voltage estimation into a single model, which reduces the error accumulation from intermediate processing steps.
The estimation accuracy of the proposed DT model is demonstrated in a real-world 84-node distribution system with raw measurements. This model showcases strong robustness against randomly missing measurements. 
A comparison with recent data-driven transformer and LSTM algorithms validates the superiority of the proposed method.

\vspace{-5 pt}
\section{Preliminary}
\subsection{Distribution System State Estimation}
 
The distribution system states are usually represented as the voltage phasors on all nodes at all existing phases. The voltages in the rectangular form are selected as the state variables, denoted as $\mathbf{x}=[\Re(\mathbf{v});\Im (\mathbf{v})]$.
% , where $\mathcal{N}:=\{0,1,2,...,N\}$ is the node set.
% and $\Xi \subset \mathcal{N} \times \mathcal{N}: =\{(i,j)|i,j\in \mathcal{N} \}$ is the edge set. 
% Indexed by 0, $\mathbf {U}_0 \in {\mathbb{C}}^{3\times1}$ denotes the slack voltage in the substation is denoted by .
% , and we define the set of other nodes as $\mathcal{N}\backslash \{0\}$. Given a multi-phase distribution network, 
% let $\Phi_i \in \{a,b,c\}$ denote a set of the existing phases on bus $i$, and the total number of the phases on all non-slack nodes is  $\mathcal{M}=\sum_{i \in \mathcal{N}\backslash \{0\}} |\Phi_i|$.
Given noisy measurements $\mathbf {z}$, the measurement equation is used to relate the states to the measurement vector, expressed by:
\begin{equation}\label{se}
 \mathbf {z}=h(\mathbf x) + \mathbf{e}
\end{equation}
where $\mathbf {e}$ is the error vector; $h(\mathbf x)$ denotes measurement functions, see the specific expression for certain measurements in \cite{9770510}.

The goal of the state estimation process is to estimate the state variables $\mathbf x$ such that the weighted sum of errors $\mathbf e$ is minimized. The WLS formulation is commonly employed to model the DSSE problem as an optimization problem:
\begin{equation} \label{se2}
\mathbf {\hat {x}} = \mathop {\mathrm {arg\,min}}\left ({\mathbf {z}- h\left (\mathbf {x}\right)}\right)^{\top}\mathbf {W}\left ({\mathbf {z}-h\left (\mathbf {x}\right)}\right)
\end{equation}
where $\mathbf {\hat {x}}$ is the estimated state vector, and $\mathbf {W}$ denotes the weight matrix that represents the user's confidence in the measured data. A widely-used choice for the weight matrix is $\mathbf{W}=\mathbf R^{-1}$, with $\mathbf R=diag({\sigma^{2}_1, \sigma^{2}_2,...,\sigma^{2}_m})$, where $\sigma^{2}_j$ represents the variance of the measurement error corresponding to the $j$th element of $\mathbf{z}$. This choice is based on two assumptions: 1) the error vector $\mathbf{e}$ has a Gaussian distribution with zero mean, and 2) the measurement errors of different elements of the measurement vector are statistically independent.
% Under these assumptions, the WLS problem transforms to the maximum likelihood estimation.

% \textit{Hybrid Measurements:} 
% Motivated by the recent prevalence of PMUs in the distribution system, t
The current engineering practice uses a hybrid metering scheme since real-time PMU sensors, which are costly, cannot solely meet the observability requirements. Moreover, the DSSE model \eqref{se} and \eqref{se2} using a hybrid metering scheme is nonlinear, making the solving process more challenging. For engineering practicality, this paper focuses on integrating hybrid voltage and power measurements from different types of sensors, typically a combination of PMUs (or/and supervisory control and data acquisition systems) and smart meters. The proposed method is also applicable to pseudo-measurements with high or unknown errors.

\subsection{Practical Data Challenges}
As mentioned before, one of the practical issues in raw measurement sets is data missing \cite{10155244, 8440475,10309251}. Distinguishing from a transmission system, which possesses high measurement redundancy, distribution feeders have lower coverage of sensors. Data missing in measurements imperatively results in the unsolvability of the DSSE model \eqref{se2}.
% Therefore, the traditional statistics-based or empirical methods are no longer applicable.  
% On the other hand, in most of the existing literature, the measurement
% missing is often expressed by a random variable obeying a
% Bernoulli distribution, where the measurements at a designated location are assumed
% to be either utterly missing or completely available. However, the assumption might be strong \cite{10155244}. 
Another challenge is the multi-modal uncertainties of measurements in distribution systems.
The heterogeneity is termed as $\textit{multi-modality}$ in computer science, which refers to the presence of multiple modes or peaks in a probability distribution \cite{baltruvsaitis2018multimodal}. Due to the stochastic nature and interdependency of renewable generation, these multi-modal uncertainties become prevalent as their penetration at the distribution level increases. Then, the assumptions used in \eqref{se2} no longer apply.
% yet it remains challenging to quantify. 
% These data challenges can significantly degrade the estimation performance of the existing methods when encountering an uncertain, arbitrary ratio of missing data. 

With the aid of AI in acquiring the missing data and mining multi-modal measurements, recent advances in DTs hold significant promise in addressing these challenges.
The AI-driven DTs can avoid the adverse impact of missing and flawed data, enabling real-time monitoring of
distribution grids and enhancing overall situational awareness,
which serves as the motivation for this paper.
\vspace{-5 pt}

\section{Proposed Interactive Attention-Based DT}
\begin{figure}[!t]
    \centering
    \hspace{1cm}
    \includegraphics[width=0.9\linewidth]{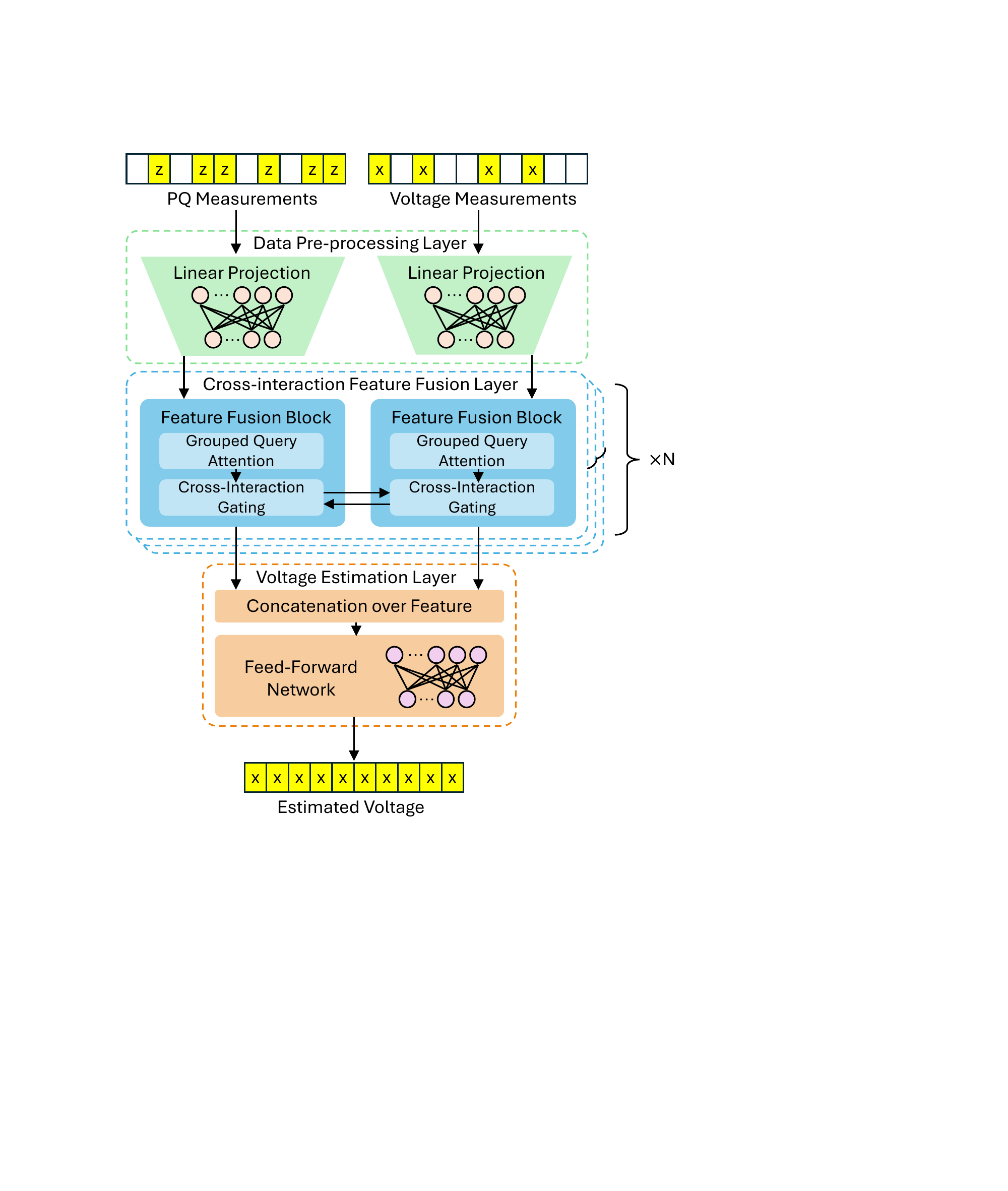}
    \caption{The proposed interactive attention-based DT architecture}
    \label{fig:overallflow}
    \vspace{-15 pt}
\end{figure}

We propose an interactive attention-based DT architecture for robust voltage estimation using time-series hybrid measurements with missing data in distribution systems. This model adopts an end-to-end framework, without data imputation, which prevents the error accumulation from separate data imputation and state estimation. Illustrated as Fig.\ref{fig:overallflow}, the proposed method consists of a data pre-processing layer, a cross-interaction feature fusion layer, and a voltage estimation layer.
Two key components in the feature fusion layer are an attention-based module for spatiotemporal feature learning and cross-interaction.
% To address the above-mentioned challenges, such as heterogeneity and the need to fuse multiple modalities of information, a deep feature fusion mechanism is proposed. 
% The model takes multiple different types of measurements and uses time-series measurements to estimate voltage states, resulting in a data-interactive DT model.
The training dataset consists of historical measurements and state records $\mathcal{D}$, denoted by $\{\mathbf{z}_t;\mathbf x_t\}_{t=0}^{t=T}$. The proposed approach estimates the voltages at time $t$ using measurement data from previous time steps, where the NN inputs are subject to measurement errors and missing. 

\vspace{-5 pt}
\subsection{Physics-Informed Data Augmentation and Transfer}
 
As the original $\mathcal{D}$ may not include sufficient scenarios with missing data, data augmentation is essential to enhance the robustness of training for DT modeling. 
To generate diverse data missing scenarios and prepare for feature extraction, we propose physics-informed data augmentation and transfer modules. These two modules consist of a physics-informed data pre-processing layer in the proposed DT.

The proposed data augmentation uses \textit{random masking} in $\mathcal{D}$ to simulate missing data during the offline training. Random masking, a widely used technique in natural language processing, designates specific positions in the input as zero during data processing. In this context, the missing probability for the $j$th measurement, denoted as $\alpha_j$, follows a Bernoulli distribution \cite{10309251}. The determination of $\alpha_j$ can be physics-informed, based on empirical knowledge or offline statistical analysis; however, this is not a prerequisite for the applicability of the proposed DT model. On the online implementation, data missing if existing is treated as zero elements in the NN input.

% Our case study demonstrates the adaptivity of this method to variations in the actual $\alpha_j$, which may differ from those used in offline training.

Due to the multimodality of hybrid measurement in the current distribution landscape, directly combining multimodal data into a single vector is less effective.
Some studies, such as \cite{9639949,NTU7913716, raghuvamsi2024distribution}, treat all modalities indiscriminately, essentially overlooking the differences in magnitudes among various physical quantities. 
In contrast, to handle power and voltage measurements in the dedicated DSSE problem, we propose a parallel linear projection method for data transfer. This approach ensures input dimension alignment of heterogeneous quantities. 
The measurements, denoted as $f_{\text{linear}}(\mathbf z_i)$, are applied separately to power and voltage measurements.
% This is because the voltage and power measurements differ in their physical properties. 
The latent representations, extracted as the output of the data pre-processing layer for each branch, are generated using the linear projection, which acts as fully connected layers without activation functions:
\begin{equation}
\label{equ:linear}
f_{\text{linear}}(\mathbf z_i) = \mathbf W_i\mathbf z_i+\mathbf b_i 
\end{equation}
% \begin{equation}
% \label{equ:linear2}
% f_{\text{linear}}(\mathbf z_s) = \mathbf W_s *\mathbf z_s+\mathbf b_s 
% \end{equation}
where $\mathbf W_i$ and $\mathbf b_i$ represents the weights of the linear projection for branch $i$. The parallel linear projection reduces high-dimensional data to a lower-dimensional form for each branch while preserving essential features in the latent space. 

The proposed data pre-processing leads to exponentially decreasing NN parameters to be learned, thereby greatly reducing the training burdens for the subsequent feature fusion.
\vspace{-9 pt}
\subsection{Cross-Interaction Feature Fusion for Voltage Estimation}

A cross-interaction feature fusion method is proposed to learn the temporal interdependencies from the latent representations of multimodal measurements for accurate state estimation. Moreover, the feature fusion layer consists of an attention-based module, followed by cross-interactions between branches.

\begin{figure}[!t]
    \centering
    \includegraphics[width=0.49\linewidth]{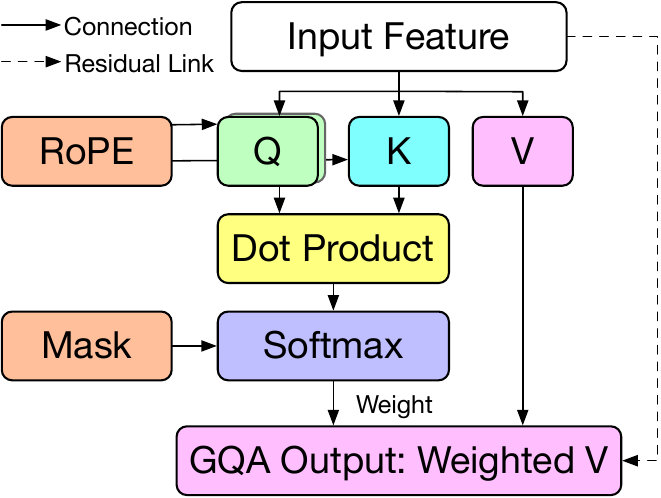}
    \caption{The proposed GQA module in the feature fusion layer}
    \label{fig:FeatureFusion}
    \vspace{-15 pt}
\end{figure}

As shown in Fig.\ref{fig:FeatureFusion}, a state-of-the-art techniques, group-queried attention (GQA) \cite{gqa}
is used to learn the temporal interdependency of feature embeddings and prevent over-fitting. The modified attention module is applied within each branch. 
Given a query $\mathbf Q$ and a set of key-value pairs
$(\mathbf K, \mathbf V)$, the classic attention mechanism computes a weighted sum
of the values based on the query and the corresponding
keys. For measurements in branch $i$, the attention mechanism calculates the weights and outputs the weighted feature vector as follows:
\begin{equation}
    \mathbf{a}_i=\text{Attention}(\mathbf Q, \mathbf K, \mathbf V) = \text{softmax}\left(\frac{\mathbf Q\mathbf K^\top}{\sqrt{d_k}}\right)\mathbf V
    \label{eq:attention}
\end{equation}
% \begin{equation}
%     \mathbf{a}_i=\mathbf w_i \mathbf V 
%     \label{eq:attentionOutput}
% \end{equation}
where $d_k$ denotes the dimension of vector $\mathbf K$; $\mathbf{a}_i \in \mathbb{R}^{T \times d}$ denotes the input features for category $i$; $T$ is the length of time steps for the measurements, and $d$ is a pre-determined feature size; softmax($\cdot$) denotes the softmax function. Building on \eqref{eq:attention}, each key in GQA is queried multiple times by multiple query subheads. Compared with the classic attention mechanism, the adopted GQA effectively reduces the number of pairwise comparisons required between queries and keys, leading to higher computational efficiency.

% Rotary positional embedding (RoPE) \cite{RoPE}, as the latest position encoding technique, is adopted for dependency modeling between elements at different positions of the time-series sequence. 
% RoPE applies positional encoding through rotations in a multi-dimensional vector space, allowing models to effectively process longer sequences. This method improves the model's ability to capture positional information without significantly increasing complexity.
% RoPE divides vectors into multiple subspaces, each rotating at different speeds, ensuring the model can accurately interpret both short-term and long-term dependencies. 
% And we can use it to train in a parallel way compared to an RNN. And the attention state table can easily process long-term relationships.

Then, a cross-interaction gating mechanism is employed to modulate information flow between branches, selectively allowing feature fusion from multi-modal measurements, while accounting for interdependency.  
The gated representation of each branch combines its own features with those modulated by the complementary branch, adaptively improving feature fusion through the following cross-interaction:
\begin{equation}
\label{eq:gate}
        \mathbf H_1^{\text{fused}} = f_{G, \mathbf \Theta_{g_2}}(\mathbf a_2) \odot \mathbf a_2 + \mathbf a_1 
\end{equation}
\begin{equation}\label{eq:gate2}
        \mathbf H_2^{\text{fused}} = f_{G, \mathbf \Theta_{g_1}}(\mathbf a_1) \odot \mathbf a_1 + \mathbf a_2
\end{equation}
where $f_{G, \Theta_{g_i}}(\mathbf{a}_i)$ represents a feed-forward NN that generates gating parameters for branch $i$ using the NN parameters $\mathbf{\Theta}_{g_i}$; $\odot$ denotes element-wise multiplication.
% \noindent $ \mathbf{H}_i^{\text{gated}} \in \mathbb{R}^{T \times d_{\text{ff}}} \gets f_{G, \mathbf{W}_{g_{j}}}(\mathbf{H}_{j}) \odot \mathbf{H}_j + \mathbf{H}_{i}$: Adaptive update in branch $i$ by selectively incorporating information from branch $j$.

% where $ \mathbf{O}_i \in \mathbb{R}^{T \times d_{\text{model}}} \gets f_{\text{linear}, \mathbf{W}_{li}}(\mathbf{H}_i^{\text{gated}})$:

% \vspace{-5 pt}
% \subsection{Loss Function and Training Process}
The output layer estimates the voltages, denoted as $\hat{\mathbf x}$, as follows:
\begin{equation}
\label{eq:v_es}
       \hat{\mathbf x} = f_{nn}(\mathbf [\mathbf O_1,\mathbf O_2])  
\end{equation}
where $[\mathbf O_1,\mathbf O_2]$ denotes the concatenation of the two branches' learned features; $\mathbf{O}_1 \in \mathbb{R}^{T \times d}= f_{\text{linear}, \mathbf{\Theta}_{l1}}(\mathbf{H}_1^{\text{fused}})$, and
$\mathbf{O}_2 \in \mathbb{R}^{T \times d}= f_{\text{linear}, \mathbf{\Theta}_{l2}}(\mathbf{H}_2^{\text{fused}})$. 

The proposed DT model is trained in an end-to-end manner and estimates the next-step voltages using a moving time window. Based on the estimated voltages over all $T$ time steps, the loss function is defined as:
\begin{equation}
    \mathcal{L} = \frac{1}{nT} \sum_{i=1}^{n} \sum_{t=1}^{T} (x_{t,i} - \hat{x}_{t,i})^2
    \label{eq:mse_loss}
\end{equation}
where $\hat{x}_{t,i}$ and $x_{t,i}$ are the estimated and the ground truth voltage from $\mathcal{D}$, respectively, for the $i$th state at time step $t$.

The proposed DT algorithm is described in the pseudo-code.

\begin{algorithm}[!h]
\caption{Proposed Attention-Interactive DT }
\label{alg:feature_fusion}
\begin{algorithmic}[1]
\STATE{\textbf{Input:} Historical dataset $\mathcal{D}:\{\mathbf{z}_t;\mathbf x_t\}_{t=0}^{t=T}$}\\
\STATE{\textbf{Parameters:} The size of hidden layers $d$ and $d_{ff}$, NN parameters $\mathbf W_i$, $\mathbf b_i$, $\mathbf{\Theta}_{gi}$, and $ \mathbf{\Theta}_{li}$ in each branch $i$, the number of the series-connected feature fusion blocks $N$}\\

\FOR{Epoch $= 1$ to $N_{ep}$}
\FOR{Epoch $t= 0$ to $T$}
% Pre-process original measurements to introduce the randomly missing data\\
\STATE \textbf{Data Pre-processing  Layer}\\
*Data Augmentation: Pre-process historical measurements at time 0 to $t$ by random input mask-out.
*Parallel Linear Projection: extract the latent representations for power and voltage measurements.
\STATE \textbf{Cross-Interaction Feature Fusion Layer} \\
* Attention-based interdependency learning\\
% 6.2 GLU activation for the embeddings $\mathbf{a}_1$ and $\mathbf{a}_2$ \\
% $\mathbf{H}_1 \in \mathbb{R}^{T \times d_{f}} \gets \phi_{\text{GLU}_{\mathbf{\Theta}_{s1}}}(\mathbf{a}_1)$\\
% $\mathbf{H}_2 \in \mathbb{R}^{T \times d_{f}} \gets \phi_{\text{GLU}_{\mathbf{\Theta}_{s2}}}(\mathbf{a}_2)$\\
* \text {Cross-interaction gating mechanism by \eqref{eq:gate} and \eqref{eq:gate2}} \\
%  $\mathbf{H}_1^{\text{fused}} \in \mathbb{R}^{T \times d_{\text{ff}}} \gets f_{\text{fused},\mathbf{W}_{g2}}(\mathbf{H}_2) \odot \mathbf{H}_2 + \mathbf{H}_1$ \\
% $\mathbf{H}_2^{\text{fused}} \in \mathbb{R}^{T \times d_{\text{ff}}} \gets f_{\text{fused},\mathbf{W}_{g1}}(\mathbf{H}_1) \odot \mathbf{H}_1 + \mathbf{H}_2$
% 6.4 \text{Linear projection to output the features} \\
% $\mathbf{O}_1 \in \mathbb{R}^{T \times d} \gets f_{\text{linear}, \mathbf{W}_{l1}}(\mathbf{H}_1^{\text{fused}})$\\
% $\mathbf{O}_2 \in \mathbb{R}^{T \times d} \gets f_{\text{linear}, \mathbf{W}_{l2}}(\mathbf{H}_2^{\text{fused}})$
\STATE \textbf{Voltage Estimation Layer} \\
Concatenate fused features to estimate voltages \eqref{eq:v_es}.\\ \STATE  Calculate the loss function via \eqref{eq:mse_loss}, and update the NN weights by stochastic gradient descent.\\
\ENDFOR
\ENDFOR
\STATE \textbf{Output:} NN parameters and $\mathbf{\hat x}_t$
\end{algorithmic}
\end{algorithm}

\vspace{-12 pt}

% \vspace{-5pt}
\section{Case Study}
The proposed DT model for state estimation is tested on a real unbalanced 84-node distribution system using raw data. The model details and historical measurements are open-sourced by the National Renewable Energy Laboratory (NREL) in \cite{oedi_2981}. Table \ref{tab:grids} provides the system details and hybrid metering scheme. The measurements are sampled every 15 minutes and include hybrid voltage magnitudes, phase angles, and power measurements. 
Note that the dataset does not disclose the statistical distribution of measurement errors for cybersecurity reasons. The dataset also provides ground-truth values of voltages, which are obtained using OpenDSS. The proposed algorithm is designed to adapt to varying data missing ratios during online inference. For demonstration, 5\% of historical measurements are randomly selected as missing data for offline training. 

The proposed model is compared with LSTM-based and transformer-based algorithms \cite{raghuvamsi2024distribution}.
% To compare the model with the existing model using RNN and the decoder-only transformer architecture \cite{decoder-only} as reference models.
For all methods, $N_{ep}=100$, with 2864 data samples used for offline training and 716 for online inference. For the transformer and the proposed model, the feature size is $d=1024$, the feedforward size is $d_{ff}=2048$, and the number of layers is $N=6$. The compared LSTM model consists of six layers, each with a hidden unit size of 1024. 
% The output consists of the real and imaginary parts of the voltage, concatenated together.
% In the proposed model, we utilize a two-branch structure, which requires the concatenation of both branches before generating the output. A two-layer FFN with residual connections is applied, containing 1024 hidden units.
The number of training epochs $N_{ep}=100$ with a learning rate of $10^{-4}$.
% and reduced by one-tenth every 30 epochs.

\begin{table}[!t]
\captionsetup{font=footnotesize}
\caption{\textsc{Summary of System and Metering Scheme}}
\label{tab:grids}
\centering
\scriptsize % Set font size to 8pt
\begin{tabular}{|c|c|c|}
\hline
\multicolumn{2}{|l|}{\multirow{2}{*}{No. Measurements/States}}    & Missing ratio\\ 
\multicolumn{2}{|l|}{} &  during offline training   \\ \hline
Power Measurements ($P$\&$Q$)& 350  & 5\%\\ \hline
Voltage phase angles &4  & 5\%  \\ \hline
Voltage Magnitudes &38  & 5\%  \\ \hline
State Variables &195 &-  \\ \hline
\end{tabular}
\vspace{-8 pt}
\end{table}

% \begin{equation}
% \text{MAE}=\frac{1}{N} \sum_{i=1}^N\left|x_i-\hat{x}_i\right|
% \end{equation}
% where $\hat{x}_i$ are the $i$th estimated state variables for $x_i$, defined by voltage magnitudes (in per unit) and phase angles (in radians) at distribution network nodes.
% $N=2n-1$ denotes the total number of estimated voltage variables, where $n$ is the number of nodes (buses).

\vspace{-5 pt}
\subsection{Estimation Accuracy and Scalability}
We investigate and compare the accuracy of the estimated voltages by the proposed method, LSTM, and transformer-based algorithms. Root mean square errors (RMSEs) and mean absolute errors (MAEs) for the estimated states, i.e., voltage magnitudes and phase angles, are calculated over multiple time steps to evaluate the overall estimation accuracy.

\begin{figure}
    \centering
    \includegraphics[width=0.95\linewidth]{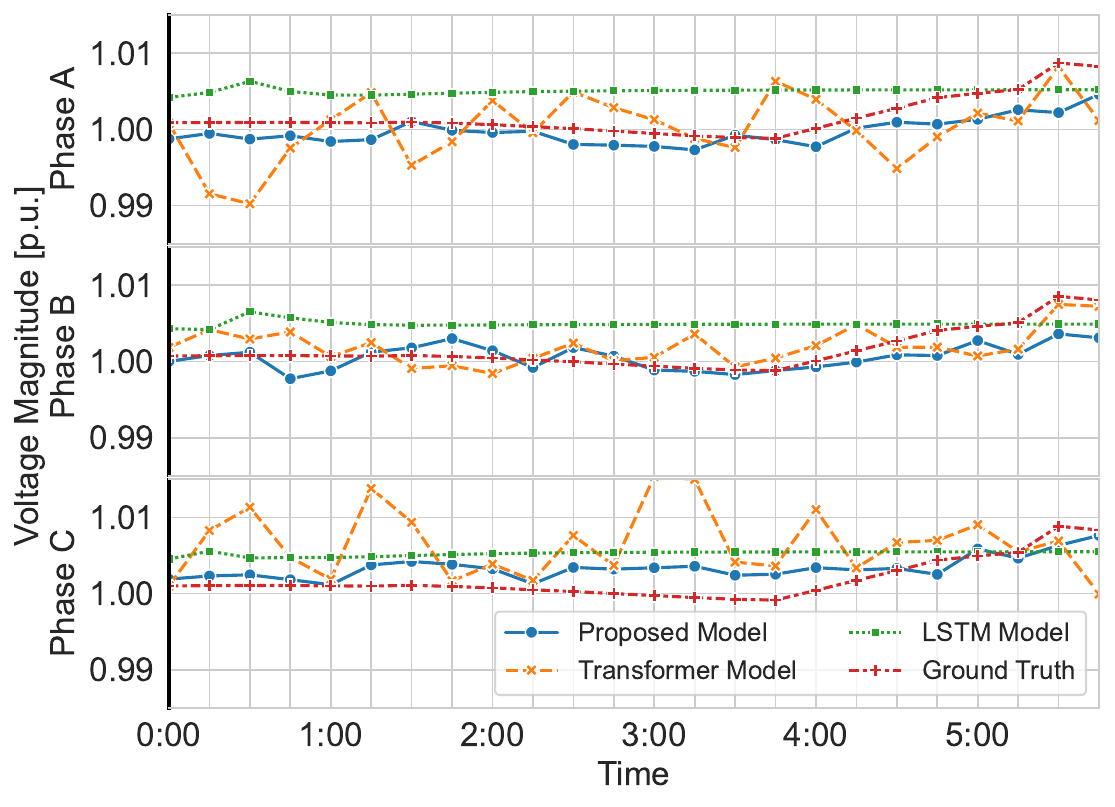}
    \vspace{-5 pt}
    \caption{Comparison of estimated voltages over multiple time steps by different methods. The ground-truth values are also plotted.}
    \label{fig:vol-mag-ot}
    \vspace{-10 pt}
\end{figure}

Fig. \ref{fig:vol-mag-ot} compares the estimated three-phase voltage magnitudes over multiple time steps at a node labeled P1UDT938LV.
It can be observed that the transformer model fails to track voltage changes under random data missing scenarios, exhibiting arbitrary fluctuations and the lowest estimation accuracy in this case. The LSTM model demonstrates apparent inertia, and the temporal voltage changes are unable to be captured due to random missing data. In contrast, the proposed model achieves the highest accuracy for voltage estimation.

\begin{figure}[!t]
    \centering
    \includegraphics[width=0.95\linewidth]{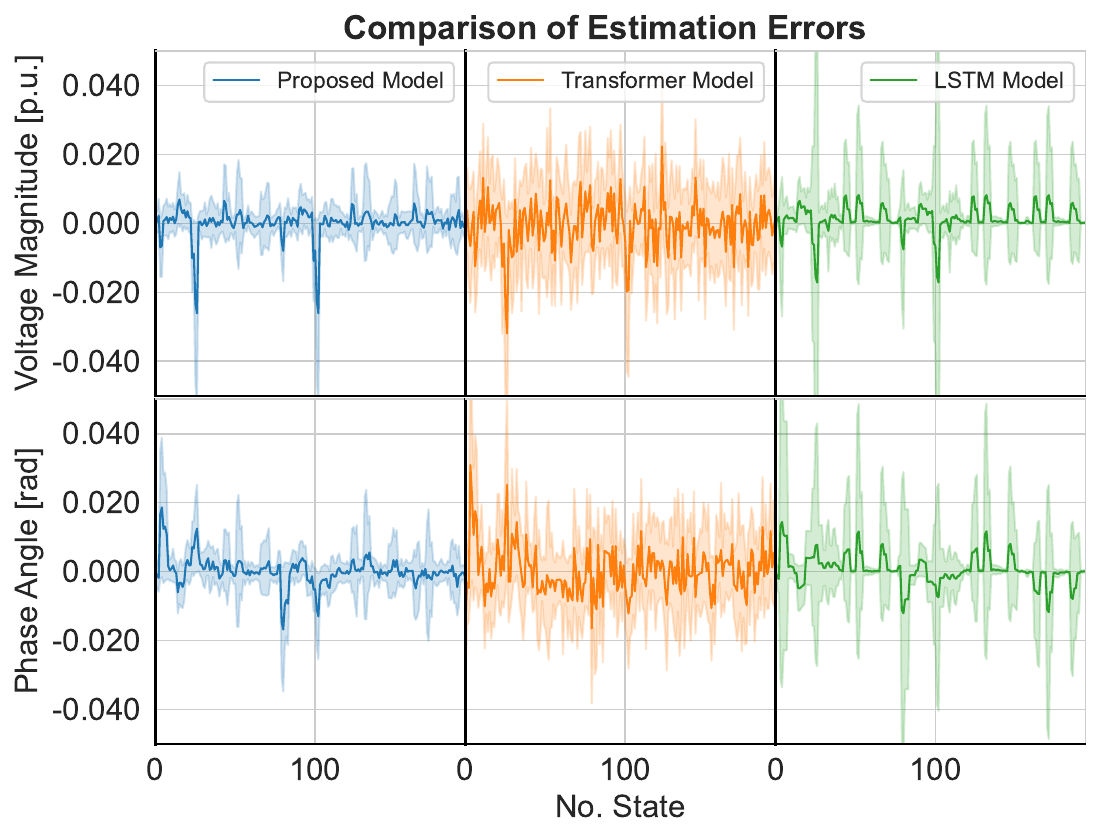}
    \caption{Comparison of the statistical distribution of errors for voltage magnitudes and phase angles in all test cases among three models.}
    \label{fig:ratios-scan}
    \vspace{-15 pt}
\end{figure}

The voltage estimation errors in terms of RMSEs and MAEs from these three methods are further compared in Table \ref{tab:rmse_mae_comparison}. The statistical distributions, depicted by the minimum, maximum, and mean errors across all test cases, are visualized in Fig.\ref{fig:ratios-scan}. It is evident that the transformer model exhibits drastic variations in estimation errors across the grid in all test cases. The test case with the highest error is produced by the LSTM model, despite relatively low mean errors overall. In contrast, the proposed DT model consistently shows stably low mean errors and the narrowest error band.

\begin{table}[!t]
\captionsetup{font=footnotesize}
\caption{\textsc{Comparison of RMSEs and MAEs}}
\label{tab:rmse_mae_comparison}
\centering
{\fontsize{8}{10}\selectfont % Set font size to 8pt with 10pt line spacing
\begin{tabular}{|c|c|c|c|}
\hline
\multirow{2}{*}{Method} & \multicolumn{2}{c|}{Voltage Magnitude} & \multicolumn{1}{c|}{Phase Angle} \\ \cline{2-4}
& RMSE [\%] & MAE [p.u.] & MAE [rad] \\ \hline
\textbf{Proposed Method} & \textbf{0.696\%} & \textbf{0.00301} & \textbf{0.00331} \\ \hline
Transformer \cite{raghuvamsi2024distribution} & 1.321\% & 0.00733 & 0.00715 \\ \hline
LSTM & 1.006\% & 0.00382 & 0.00433 \\ \hline
\end{tabular}
}
\vspace{-10 pt}
\end{table}

\vspace{-5pt}
\subsection{Robustness against Random Data Missing}

To demonstrate the adaptivity of the proposed method to variations in the ratio of random missing data between the training and testing phases, the missing data ratio $\alpha$ is dynamically varied during online inference. The proposed DT model is trained with $\alpha=5\%$, while
$\alpha$ ranges from 0\% to 40\% during the online testing to investigate the impact of random data missing. 

\begin{figure}[!t]
    \centering
    \includegraphics[width=0.9\linewidth]{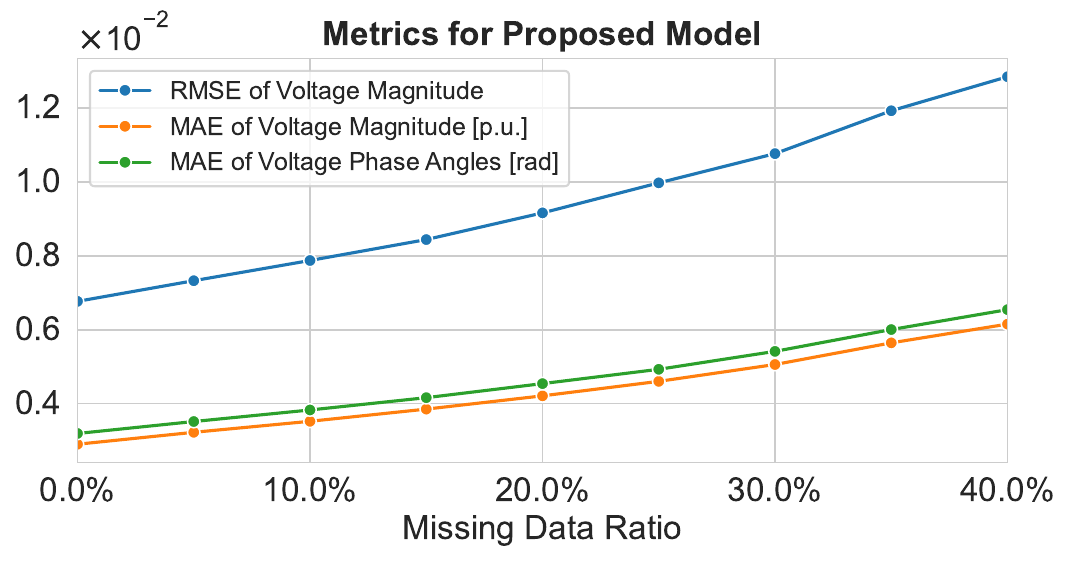}
    \caption{Error trends of the proposed method in RMSEs and MAEs under different data missing ratios}
    \label{fig:ratios-scan2}
    \vspace{-17 pt}
\end{figure}

The effect of different data missing ratios on estimation errors is illustrated in Fig. \ref{fig:ratios-scan2}.
The results show that the proposed algorithm adapts effectively to a broad range of missing data ratios, even when different from the training scenario. Although the estimation errors increase with higher missing ratios, the errors remain acceptable. Specifically, with 40\% of the measurements randomly missing, the MAEs are at most 0.00606 p.u. for voltage magnitudes and 0.00628 rad for voltage phase angles. These results demonstrate the robustness of the proposed method against random data missing.

\vspace{-5 pt}
\section{Conclusion}
This paper proposes a novel interactive attention-based DT architecture for DSSE under random data missing scenarios. Unlike data-driven
methods that focus solely on virtual modeling, the proposed DT model
integrates physical knowledge into the NN design.
% The proposed model proposes a state-of-the-art attention-based feature extraction and a novel deep cross-interaction feature fusion for robust state estimation. 
Unlike traditional model-based methods that rely on uncertainty-specific assumptions,
the proposed model offers a robust DT solution that accommodates random and arbitrary portions of missing data, up to 40\% of total measurements. 
This level of missing data ratio and randomness is commonly observed in field data, highlighting the method's applicability to real-world grids.
% The adaptivity and scalability of the proposed DT model are validated on a real-world unbalanced 84-bus distribution system with random, arbitrary portions (up to 40\% of total measurements) of data missing.
% This paper unlocks the potential of DT models for dedicated physical model-free state estimation applications in unobservable distribution grids.

\IEEEpubidadjcol

\bibliographystyle{IEEEtran}
\bibliography{IEEEabrv,Citation}
\let\mybibitem\bibitem

\end{document}